\documentclass[aps,prd,twocolumn,showpacs,floatfix,preprintnumbers,amsmath,amssymb,nofootinbib,groupedaddress, 
superscriptaddress, notitlepage]{revtex4-1}

\input epsf
\usepackage{graphicx}
\usepackage{color}
\usepackage{mathtools}
\usepackage{mathbbol}
\usepackage[utf8]{inputenc}
\usepackage[normalem]{ulem}
\usepackage[dvipsnames]{xcolor}
\usepackage[colorlinks=true,citecolor=blue,linkcolor=blue,urlcolor=blue, backref=false,pdfborder={0 0 0}]{hyperref}
\usepackage{float}

\newcommand{\beq}{\begin{equation}}
\newcommand{\eeq}{\end{equation}}
\newcommand{\barr}{\begin{eqnarray}}
\newcommand{\earr}{\end{eqnarray}}
\newcommand{\dbk}{DBK13}
\newcommand{\rhobar}{\ensuremath{\overline{\rho}}}

\newcommand{\bs}{\boldsymbol}

\newcommand{\physrep}{Physics Reports}
\newcommand{\jcap}{Journal of Cosmology and Astroparticle Physics}

\begin{document}
\title{Exact treatment of weak dark matter-baryon scattering for linear-cosmology observables}

\author{Yacine Ali-Ha\"imoud}
\affiliation{Center for Cosmology and Particle Physics, Department of Physics, New York University, New York, New York 10003, USA} 
\author{Suroor Seher Gandhi}
\affiliation{Center for Cosmology and Particle Physics, Department of Physics, New York University, New York, New York 10003, USA} 
\author{Tristan L. Smith}
\affiliation{Department of Physics and Astronomy, Swarthmore College, 500 College Ave., Swarthmore, PA 19081, United States}

\begin{abstract}

Elastic scattering of dark matter (DM) particles with baryons induce cosmological signals that
may be detectable with modern or future telescopes. For DM-baryon scattering cross sections scaling with negative powers of relative velocity, $\sigma_{\chi b}(v) \propto v^{-2}, v^{-4}$, such interactions introduce a momentum-exchange rate that is nonlinear in DM-baryon bulk relative velocities, thus not amenable for inclusion as-is into standard linear cosmological Boltzmann codes. Linear ansatzes have been adopted in past works, but their accuracy is unknown as they do not arise from first-principles derivations. In this work, for the first time, we construct a rigorous framework for computing linear-cosmology observables as a perturbative expansion in $\sigma_{\chi b}$. We argue that this approach is accurate for Cosmic Microwave Background (CMB) angular power spectra when most or all of the DM is scattering with baryons with cross section $\sigma_{\chi b}(v) \propto v^{-2}, v^{-4}$. We derive exact formal expressions for CMB power spectra at linear order in $\sigma_{\chi b}$, and show that they only depend on a specific velocity integral of the momentum-exchange rate. Consequently, we can obtain the exact power spectra at linear order in $\sigma_{\chi b}$ by substituting the original nonlinear momentum-exchange rate with a uniquely specified linear rate. Serendipitously, we find that the exact substitution we derive from first principles precisely coincides with the most widely used linear ansatz, thus placing previous CMB-anisotropy upper bounds on more solid footing. In addition to finally providing an exact cosmological solution to the DM-baryon scattering problem in a well-defined region of parameter space, the framework we construct opens the way to computing higher-order correlation functions, beyond power spectra, which are promising yet unexplored probes of DM-baryon scattering.
\end{abstract}

\maketitle

\section{Introduction}
%
As it stands today, we do not understand the nature of dark energy nor dark matter (DM) in the $\Lambda$CDM model, and they act as placeholders for roughly 95\% of the Universe's energy budget. Fixing DM to be ``cold" (i.e.~with negligible velocity dispersion) is a highly generic constraint that describes observations exceedingly well. This general property is theoretically consistent with a myriad of DM models, see e.g.~Refs.~\cite{Green:2022review, Arbey:2021review} for recent reviews. Given the rich diversity of phenomena that we can now observe, cosmology and astrophysics offer some of the most promising prospects of distinguishing the true nature of DM amidst the many theoretical possibilities consistent with the cold DM paradigm  \cite{Drlica-Wagner:2022review}.

In this work, we focus on a generic class of particle DM models which include scattering interactions with Standard Model particles, in particular baryons\footnote{In the literature on scattering interactions between DM and Standard Model (SM) particles, any massive SM scatterer is generically referred to as a \textit{baryon}, and we adhere to the same terminology in this work.}. Such models are the primary target of direct-detection experiments, but may also be tested through a variety of astrophysical and cosmological signatures \cite{Gluscevic:2019yal, Boddy:2022sss}. Cosmic probes of DM-baryon scattering include spectral distortions of the cosmic microwave background (CMB) \cite{Ali-Haimoud:2015pwa,Ali-Haimoud:2021lka}, the global average and fluctuations of the redshifted 21-cm signal from the cosmic dark ages \cite{ Tashiro:2014_21cm, Munoz:2015bca} and cosmic dawn \cite{Kovetz:2018_21cm, Slatyer:2018_21cm, Driskell:2022_21cm}, CMB temperature and polarization anisotropies \cite{Chen:2002lss, Dvorkin:2013cea, Gluscevic:2018cmb, Boddy:2018kfv, Boddy:2018wzy, Xu:2018, Li:2018velindep, Li:2023act_dr4}, the Lyman-$\alpha$ forest \cite{Dvorkin:2013cea, Xu:2018, Rogers:2022lss}, and Milky-Way satellite count \cite{Buen-Abad:2022mw, Bechtol:2022sss, Maamari:2020aqz, Nadler:2019zrb}. 

Finding evidence of subtle signatures of DM interactions hinges on the robustness and accuracy of the theoretical framework used to predict them. Almost all cosmological tests of DM-baryon scattering rely on \emph{linear} cosmological perturbation theory, whether directly (e.g.~for CMB anisotropies), or as an input for further nonlinear computations (e.g.~for Milky-Way satellite count). However, DM-baryon scattering introduces into the baryon and DM equations of motion a momentum-exchange term that is intrinsically nonlinear in the DM and baryon bulk velocities. This nonlinearity renders the standard linear framework of solving for the evolution of cosmological perturbations inapplicable \cite{Dvorkin:2013cea, Boddy:2018wzy}. 

A decade ago, \citet{Dvorkin:2013cea} (DBK13 hereafter) suggested a simple linearized ansatz to circumvent this issue, consisting in incorporating bulk relative motions as an effective global increase to thermal relative motions. As acknowledged by DBK13, this ``mean-field" approach should, a priori, only be valid on lengthscales much larger than those on which bulk relative velocities vary significantly. In practice, however, relative motions fluctuate on scales as large as the sound horizon at photon-baryon decoupling \cite{Tseliakhovich:2010bj}, that is, lengthscales comparable to, or even larger than, those relevant to cosmological probes. As noted by DBK13, probes of DM-baryon interactions with cross sections $\sigma_{\chi b}(v) \propto v^n$ with $n \geq 0$ are mostly sensitive to epochs at which bulk relative velocities are small in comparison to their thermal counterparts, hence the DBK13 ansatz (or, even more simply, neglecting bulk relative motions in the momentum-exchange coefficient altogether) should be reasonably accurate in these cases. However, for cross sections scaling as negative powers of relative velocity, $\sigma_{\chi b}(v) \propto v^{-2}, v^{-4}$, the accuracy of the DBK13 approximation has remained unknown until now. Despite this caveat, the DBK13 ansatz, or more sophisticated but closely related versions of it \cite{Boddy:2018wzy}, have been the standard approach in most analyses of cosmological data to constrain DM-baryon scattering.  

In this work, we formulate a novel perturbative approach that yields the exact CMB power spectra for sufficiently weak DM-baryon interactions, and is especially applicable for DM-baryon cross sections $\sigma_{\chi b}(v) \propto v^{-2}, v^{-4}$, if all or most of the DM interacts with baryons. This is the first implementation of DM-baryon scattering in the context of linear cosmology that is exact in a well-defined and physically-relevant region of parameter space.
Our formulation is novel in that it treats the interaction cross-section as the small parameter of interest, while allowing to keep the full nonlinear dependence on cosmological velocity fields. Using this framework, we derive an exact expression for CMB power spectra at linear order in the DM-baryon cross section, which we show to only depend on a specific velocity integral of the nonlinear momentum-exchange rate. As a consequence, we derive the unique substitution for the nonlinear DM-baryon momentum-exchange rate that is linear in bulk velocities, yet leads to the exact CMB power spectra at linear order in the cross section. 
Unexpectedly, we discover that the widely used \dbk\ approach, which was originally presented as an approximate phenomenological workaround, is in fact precisely equal to the unique linear substitution that we derive rigorously, starting from a completely different perspective. In addition to developing a novel and rigorous methodology, our work thus also places existing constraints relying on the DBK13 approach on a much firmer footing. 

The remainder of this paper is organized as follows. In Sec.~\ref{sec:general}, we review the general equations governing momentum and heat exchanges sourced by DM-baryon scattering. In Sec.~\ref{sec:nonlinear}, we describe the nonlinearity problem, and the standard phenomenological solutions that have been proposed in the literature. The expert reader may skip directly to Sec.~\ref{sec:solution}, where we describe our approximation scheme and derive an exact solution in the weak-interaction limit. We conclude and describe future work in Sec.~\ref{sec:conclusion}. Appendix \ref{app:constrained} provides the proof for an intermediate result on constrained averages that is relevant to the exact solution presented in Sec.~\ref{sec:solution}.

\section{General equations}\label{sec:general}

We consider a particle $\chi$ of mass $m_\chi$, making up all or part of the DM, and scattering elastically with baryons with a velocity-dependent momentum-transfer cross section $\sigma_{\chi b}(v)$. In order to keep the discussion simple, we consider the case where $\chi$ scatters with only one type of ``baryons" (which could be hydrogen or helium nuclei or atoms, or electrons), with mass $m_b$. We explain in Sec.~\ref{sec:multiple-scat} how our results carry over when $\chi$ scatters simultaneously with multiple baryonic components.

Due to their frequent self-interactions, baryons are described at all relevant times and at any location by a non-relativistic Maxwell-Boltzmann (MB) velocity distribution with mean velocity $\bs{V}_b$ and temperature $T_b$. As pointed out in Ref.~\cite{Ali-Haimoud:2018dvo}, the particle $\chi$ need not have a MB distribution at all times if it has weak self-interactions. However, extrapolating from the results of Refs.~\cite{Ali-Haimoud:2018dvo, Gandhi:2022tmt}, approximating $\chi$'s velocity distribution by a MB distribution should be accurate at the order-unity level. Given that we are focused on a separate aspect of the problem in this work, and given the tremendous simplifications afforded by the MB approximation, we shall make the common assumption that $\chi$ is also described at all times by a non-relativistic MB velocity distribution, with mean velocity $\bs{V}_\chi$ and temperature $T_\chi$. Note that this is an accurate description if $\chi$ self-interacts frequently enough (i.e.~more than once per Hubble time), hence our results are accurate in that limit. As a consequence, the distribution of \emph{relative} velocities between $\chi$ and baryons is also a MB distribution, with mean $\bs{V}_{\chi b} \equiv \bs{V}_\chi - \bs{V}_b$, and variance per axis 
\beq
(T/m)_{\chi b} \equiv T_\chi/m_\chi + T_b/m_b. 
\eeq
Note that the variables $\bs{V}_b, \bs{V}_\chi, T_b$ and $T_\chi$ are all time- and space-dependent.  

We consider specifically the high-redshift Universe, when cosmological inhomogeneities are small, thus justifying the use of perturbation theory in the standard (no DM-baryon scattering) case. Elastic scattering between the particle $\chi$ and baryons results in two effects, taking a particularly simple form within the MB approximation:

(\textit{i}) An exchange of momentum between baryons and $\chi$, as well as new pressure force for $\chi$, expressed as additional terms in the baryons' and $\chi$'s momentum equations:
\barr
\dot{\bs{V}}_\chi &=& \dot{\bs{V}}_\chi\big{|}_{\rm std}  - \overline{\rho}_b \Gamma_V \bs{V}_{\chi b} - \frac1{\overline{\rho}_\chi} \bs{\nabla} P_\chi , \label{eq:dotVchi}\\
\dot{\bs{V}}_b &=& \dot{\bs{V}}_b\big{|}_{\rm std} +\overline{\rho}_\chi  \Gamma_V \bs{V}_{\chi b} \label{eq:dotVb},
\earr
where overdots denote time derivatives\footnote{We use regular time rather than conformal time to keep expressions most compact. This implies that gradients are proper rather than comoving, so that that $\bs{\nabla} \rightarrow a^{-1} \bs{k}$ in Fourier space, where $a$ is the scale factor and $\bs{k}$ is the comoving wavenumber.}, the subscript ``std" labels the ``standard" terms, present even without DM-baryon interactions, $\overline{\rho}_b$ and $\overline{\rho}_\chi$ are the average mass densities of baryons and $\chi$, respectively, and the coefficient $\Gamma_V$ will be described shortly. Note that, with or without DM-baryon scattering, one may neglect the small density fluctuations of baryons and DM in the above equations at the times of interest. 

(\textit{ii}) An exchange of heat between baryons and $\chi$, which would otherwise remain cold, as well as a dissipation of relative velocities into heating both baryons and $\chi$ \cite{Munoz:2015bca}:
\barr
\dot{T}_\chi &=& \dot{T}_\chi\big{|}_{\rm std}+ n_b \Gamma_T(T_b - T_\chi) + \frac23 \overline{\rho}_b \mu_{\chi b} \Gamma_V V_{\chi b}^2, \label{eq:Tchidot}\\
\dot{T}_b &=& \dot{T}_b\big{|}_{\rm std} + n_\chi \Gamma_T(T_\chi - T_b) +  \frac23 \overline{\rho}_\chi \mu_{\chi b} \Gamma_V V_{\chi b}^2, \label{eq:Tbdot}
\earr 
where  $\mu_{\chi b} \equiv m_\chi m_b/(m_\chi + m_b)$ is the DM-baryon reduced mass, and $n_b = \rho_b/m_b$ and $n_\chi = \rho_\chi/m_\chi$ are the number densities of baryons and $\chi$, respectively, which may be approximated by their spatial averages $\overline{n}_b$ and $\overline{n}_\chi$ in these equations for the cosmological probes of interest. The terms proportional to the coefficient $\Gamma_T$ represent ``thermal" heat exchange, while the terms\footnote{Equations \eqref{eq:Tchidot} and \eqref{eq:Tbdot} can also be derived by substituting $T_\chi = \mu_{\chi b} (T/m)_{\chi b} + \frac{m_\chi}{m_b + m_\chi}(T_\chi - T_b)$ in the second term of Eq.~(59) in Ref.~\cite{Ali-Haimoud:2018dvo}} proportional to $\Gamma_V V_{\chi b}^2$ account for the dissipation of bulk relative velocities into heat \cite{Munoz:2015bca}. 

The coefficients $\Gamma_V$ and $\Gamma_T$ are both obtained by taking a weighted integral of the momentum-transfer cross section $\sigma_{\chi b}(v)$ over the distribution of $\chi$-baryon relative velocities. Under the MB approximation, they are functions of $V_{\chi b}$ and $(T/m)_{\chi b}$ alone. The specific functional forms of $\Gamma_V\left(V_{\chi b}, (T/m)_{\chi b}\right)$ and $\Gamma_T\left(V_{\chi b}, (T/m)_{\chi b}\right)$ depend on the velocity dependence of the cross section. For instance, for power-law cross sections $\sigma_{\chi b}(v) \propto v^n$, these coefficients can be expressed in terms of hypergeometric functions \cite{Munoz:2015bca, Boddy:2018kfv}. 

\section{The nonlinearity problem and existing phenomenological solutions}\label{sec:nonlinear}

It is well known that bulk relative velocities $V_{\chi b}$ may be comparable to, or even a few times larger than, their thermal counterpart $(T/m)_{\chi b}$ \cite{Tseliakhovich:2010bj, Dvorkin:2013cea}. As a consequence, in general they cannot be neglected inside the rates $\Gamma_V$ and $\Gamma_T$, which thus depend \emph{non-perturbatively} on the \emph{local} bulk relative velocity $V_{\chi b}(t, \bs{x})$. This renders the momentum equations intrinsically nonlinear, even if baryon and DM perturbations remain small for the observables of interest. Besides this obvious nonlinearity, the momentum equations also depend on temperatures, whose evolution in turn depends nonlinearly on relative velocities, through $\Gamma_T$ and the dissipation term. 

These nonlinearities can be accurately accounted for in the context of 21-cm tomography \cite{Munoz:2015bca}, using the fact that relative velocities are coherent on sub-Mpc scales \cite{Tseliakhovich:2010bj}, and that baryon velocities are no longer affected by photon drag at the relevant epochs, which significantly simplifies the problem. However, in the context of CMB anisotropies or other large-scale-structure probes, which are sensitive to DM-baryon interactions around and prior to photon-baryon decoupling, these nonlinearities render the problem not amenable for inclusion as-is into cosmological linear Boltzmann codes. Note that, due to the different time dependence of bulk and thermal relative velocities, it is well known that the nonlinearity problem is most pronounced for cross sections which decrease steeply with velocity, such as Coulomb-like cross sections $\sigma_{\chi b}(v) \propto v^{-4}$ and dipole-charge cross sections $\sigma_{\chi b} \propto v^{-2}$, see Refs.~\cite{Dvorkin:2013cea, Boddy:2018wzy} for more detailed discussions on this point.\\

In order to bypass this nonlinearity issue, a series of assumptions are usually made in the literature, without proper justification:\\

(\textit{i}) It is assumed that, for scalar adiabatic initial conditions, DM and baryon velocities remain curl-free even with DM-baryon scattering, despite the nonlinear momentum-exchange term, which is bound to source curls. As a consequence, $\chi$ and baryon velocity fields are assumed to be fully described by their divergences, $\theta_\chi$ and $\theta_b$, respectively. \\

(\textit{ii}) The velocity divergences are assumed to satisfy the following equations:
\barr
\dot{\theta}_\chi &=& \dot{\theta}_\chi\big{|}_{\rm std} + \overline{\rho}_b \Gamma_V (\theta_b - \theta_\chi) - \frac1{\overline{\rho}_\chi}\nabla^2 P_\chi, \label{eq:dottheta_chi}\\
\dot{\theta}_b &=& \dot{\theta}_b\big{|}_{\rm std} + \overline{\rho}_\chi \Gamma_V (\theta_\chi - \theta_b), \label{eq:dottheta_b}
\earr
where $\Gamma_V$ is still assumed to be a function of the local $V_{\chi b}$. These equations do not properly account for the spatial dependence of $\Gamma_V$ through its dependence on $V_{\chi b}$. Indeed, correctly taking the divergence of Eq.~\eqref{eq:dotVb} gives
\beq
\dot{\theta}_b = \dot{\theta}_b\big{|}_{\rm std} + \overline{\rho}_\chi \Gamma_V (\theta_\chi - \theta_b) + \overline{\rho}_\chi \bs{\nabla} \Gamma_V \cdot \bs{V}_{\chi b}, 
\eeq
and similarly for Eq.~\eqref{eq:dotVchi}. Note that approximations (\textit{i}) and (\textit{ii}) do not deal with the nonlinearity issue \emph{per se}, but allow to start from a system of equations closer to the one usually solved absent DM-baryon scattering.\\

(\textit{iii}) In order to get rid of the nonlinear dependence of the momentum-exchange rate on $V_{\chi b}$, the most commonly used approach, presented in DBK13, consists in making the following ``mean-field" substitution :
\beq
\Gamma_V(V_{\chi b}; (T/m)_{\chi b}) \rightarrow \Gamma_V\left(0; (T/m)_{\chi b} + \frac13 \langle V_{\chi b}^2 \rangle\right), \label{eq:Dvorkin}
\eeq
and similarly for the heat-exchange rate $\Gamma_T$, where $\langle V_{\chi b}^2 \rangle$ is the time-dependent but spatially-homogeneous variance of bulk relative velocities. This substitution consists in including bulk relative motions as effective additional thermal motions. In addition, the dissipation of relative velocities into heat is typically neglected, meaning that the terms proportional to $\Gamma_V V_{\chi b}^2$ are neglected in the temperature evolution equations. Lastly, $T_b$ and $T_\chi$ are assumed to be spatially homogeneous.\\

(\textit{iv}) Even with these substitutions, the set of equations to be solved remain nonlinear through the dependence of $\Gamma_V$ and $\Gamma_T$ on $\langle V_{\chi b}^2 \rangle$. The most widely used approximation thus consists in using the standard value of $\langle V_{\chi b}^2 \rangle$, obtained when DM-baryon interactions are neglected:
\beq
\langle V_{\chi b}^2 \rangle \rightarrow \langle V_{\chi b}^2 \rangle^{\rm std}.
\eeq

A more sophisticated approach was proposed in Ref.~\cite{Boddy:2018wzy}, hereafter BG+18. For a given comoving Fourier mode $k$, the authors split the variance of bulk relative velocities into two pieces: a large-scale piece $V_{\rm flow}^2(k)$ resulting from perturbations with wavenumbers smaller than $k$, and a small-scale piece $V_{\rm rms}^2(k)$ resulting from perturbations with wavenumbers larger than $k$, such that $V_{\rm flow}^2(k) + V_{\rm rms}^2(k) = \langle V_{\chi b}^2 \rangle$ for any $k$. Instead of Eq.~\eqref{eq:Dvorkin}, BG+18 adopt the following $k$-dependent substitution in the momentum equations
\barr
&&\Gamma_V(V_{\chi b}; (T/m)_{\chi b}) \nonumber\\
&&\rightarrow \Gamma_V\left(V_{\rm flow}(k); (T/m)_{\chi b} + \frac13 V_{\rm rms}^2(k) \right), \label{eq:Boddy}
\earr
and use $V_{\chi b} \rightarrow \langle V_{\chi b}^2 \rangle^{1/2}$ throughout in the heat equation, thus approximately including the dissipation of relative velocities into heat, but still assuming homogeneous temperatures. Within this setup, the authors of BG+18 do account for the feedback of DM-baryon interactions on relative velocities. They do so iteratively by starting from the standard values (i.e.~without DM-baryon scattering) of $V_{\rm flow}$ and $V_{\rm rms}$, solving for the evolution of perturbations given these values, and updating the power spectrum of relative velocities (assumed to remain Gaussian-distributed), hence $V_{\rm flow}$ and $V_{\rm rms}$, and so on, until convergence. 

Interestingly, when $\chi$ makes up all of the DM, BG+18 find only small differences between their implementation and that of DBK13, even for a cross section $\sigma_{\chi b}(v) \propto v^{-4}$, for which relative velocities are expected to be most relevant \cite{Dvorkin:2013cea}. This can be understood a posteriori for the following reasons. First, one can show that, over the entire expected range of bulk-to-thermal velocity ratios $\langle V_{\chi b}^2 \rangle^{1/2}/ (T/m)_{\chi b}^{1/2} \leq 3$ \cite{Dvorkin:2013cea}, and for \emph{any} values of $V_{\rm rms}$ and $V_{\rm flow}$, the phenomenological $\Gamma_V$ coefficients given in the right-hand-sides of Eqs.~\eqref{eq:Dvorkin} and \eqref{eq:Boddy} differ by no more than 20\%. Second, in the case where $\chi$ makes up all of the DM, relative velocities are only affected perturbatively by DM-baryon scattering \cite{Boddy:2018wzy}, implying that the more sophisticated iterative approach of BG+18 does not lead to significant corrections. 

The close numerical agreement of the standard approach of DBK13 and of the more sophisticated implementation of BG+18 should not be taken as a reinforcement of the accuracy of either method. First, neither of these approaches is justified from first principles, nor arises from a well-defined perturbative scheme. Second, other equally intuitive ansatzes could have been adopted, with significant differences in the effective coefficient $\Gamma_V$, thus in the resulting limits on the DM-baryon cross section. For example, one could have chosen to substitute $\Gamma_V(V_{\chi b})$ by its cosmological mean $\langle \Gamma_V(V_{\chi b}) \rangle$, obtained by taking its average over the distribution of relative velocities. We find that for $\sigma_{\chi b}(v) \propto v^{-4}$, the velocity-averaged $\Gamma_V$ may be up to \emph{twice} the phenomenological $\Gamma_V$ prescribed by DBK13. Such order-unity differences between equally-sensible choices imply that one should consider the current implementations of DM-baryon scattering and the resulting upper limits as no more accurate than the order-unity level, unless a comparison to an exact result shows otherwise.

\section{A rigorous solution in the weak-interaction regime}\label{sec:solution}

We now describe the core of our new work, namely deriving an \emph{exact} solution for linear-cosmology observables in the limit of weak DM-baryon interactions. 

\subsection{Defining the regime of interest}\label{sec:regimeOfInterest}

We consider the regime in which the effect of DM-baryon scattering on a given observable is quasi-linear in the cross section $\sigma_{\chi b}$, seen as a \emph{parameter} (more precisely, the parameter would be the cross-section evaluated at some characteristic velocity). Note that this notion of linearity in $\sigma_{\chi b}$ is distinct from the (non)linearity of the fluid equations in their \emph{variables} discussed earlier, which persists even for arbitrarily small $\sigma_{\chi b}$. Conversely, even with the linearization ansatz proposed by DBK13, which renders the fluid equations linear in all their variables, the effect of DM-baryon interactions could still be non-perturbative, hence nonlinear in $\sigma_{\chi b}$, for sufficiently large cross sections. 

Explicitly, we expect the the effect of DM-baryon interactions to be quasi-linear in $\sigma_{\chi b}$ for CMB anisotropies if all or most of the DM is interacting with a cross section $\sigma_{\chi b}(v) \propto v^{-4}$. Indeed, in this case, DM and baryons are initially decoupled, and eventually become coupled. This coupling must happen sufficiently late after the last-scattering epoch, else it would strongly perturb CMB anisotropies, which are consistent with being close to the standard $\Lambda$CDM expectation. Hence, at the times relevant to CMB anisotropies, DM-baryon scattering must be perturbative in $\sigma_{\chi b}(v) \propto v^{-4}$, if a significant fraction of the DM is interacting with baryons.

\begin{figure*}[h]
    \centering
    \includegraphics[scale=0.6, trim={.0cm .365cm .0cm .0cm}, clip]{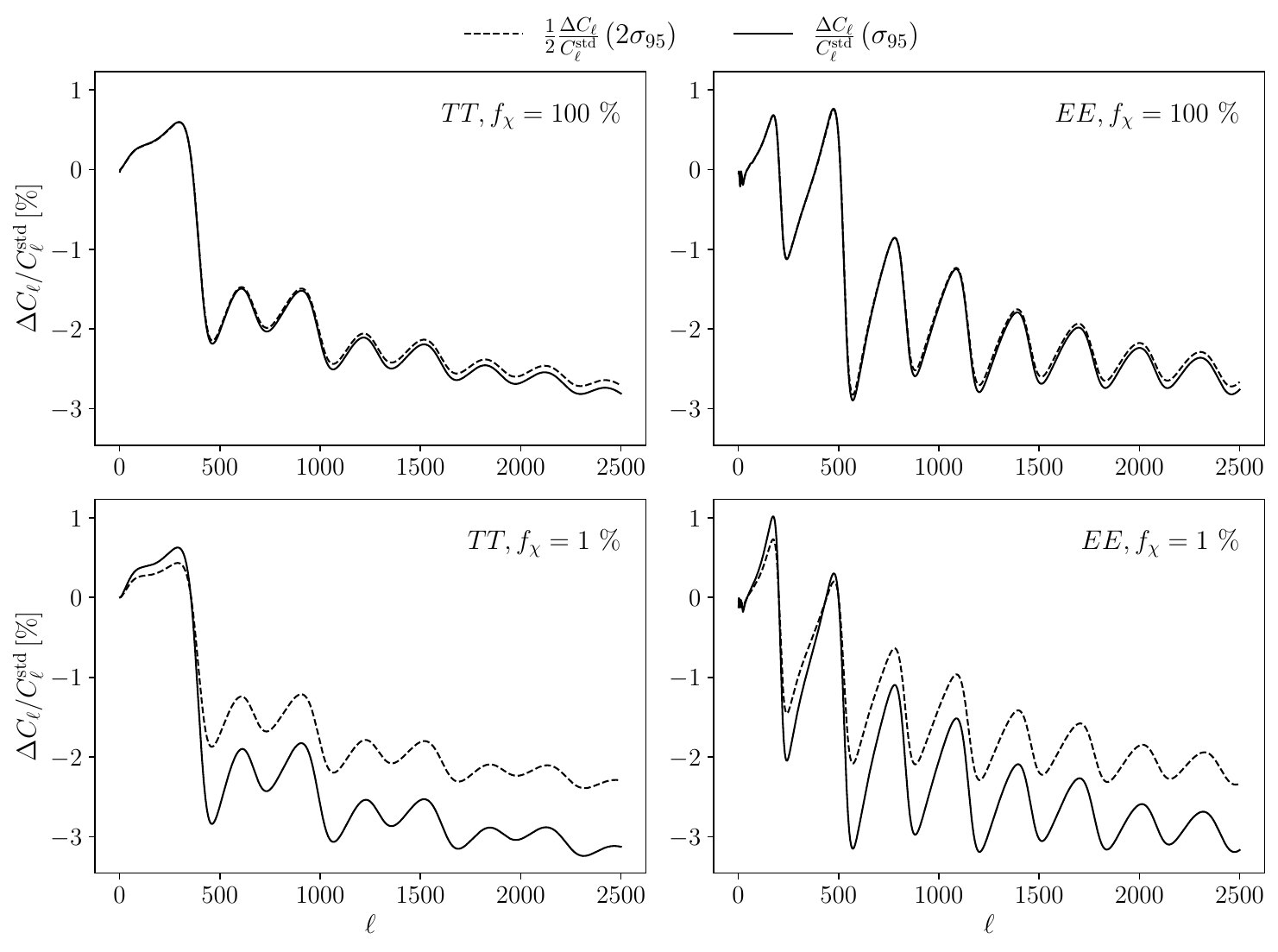}
    \vspace{-0.35cm}
    \caption{Fractional change in CMB temperature (left column) and $E$-mode polarization (right column) lensed angular power spectra, due to DM-baryon interactions, computed with \textsc{class} \cite{Blas:2011rf, Becker:2020hzj}. Solid lines show the effect of a DM-baryon cross section saturating the CMB-anisotropy 95\% upper limits of Ref.~\cite{Boddy:2018wzy}, and dashed lines show half the effect of twice that cross section. The two lines should overlap when CMB power spectra are linear in the cross section. We see that this is very nearly the case when all the DM is interacting with baryons (i.e., $f_\chi = 100\%$, in the upper panels), but that CMB power spectra are significantly nonlinear in the cross section when only one percent of the DM is interacting ($f_\chi = 1\%$, in the lower panels). These power spectra are computed specifically for a fixed DM particle mass $m_\chi = 1$ MeV interacting with all baryons with a Coulomb-like cross section $\sigma_{\chi b}(v) \propto  v^{-4}$, but we have explicitly checked that the same qualitative conclusions apply to any DM mass MeV $\leq m_\chi \leq$ GeV, and also for $\sigma_{\chi b}(v) \propto v^{-2}$. Specifically, we used $\sigma_{\chi b}(v) = \sigma_{95} v^{-4}$, with $\sigma_{95} = 1.7\times 10^{-41}$ cm$^2$ and $5.5\times 10^{-39} {\rm cm}^2$ for $f_\chi = 100\%$ and $1\%$ respectively.}
    \label{fig:DCl}
\vspace{.15cm}
    \includegraphics[scale=0.7, trim={2.12cm 6.81cm 2.13cm 6.8cm}, clip]{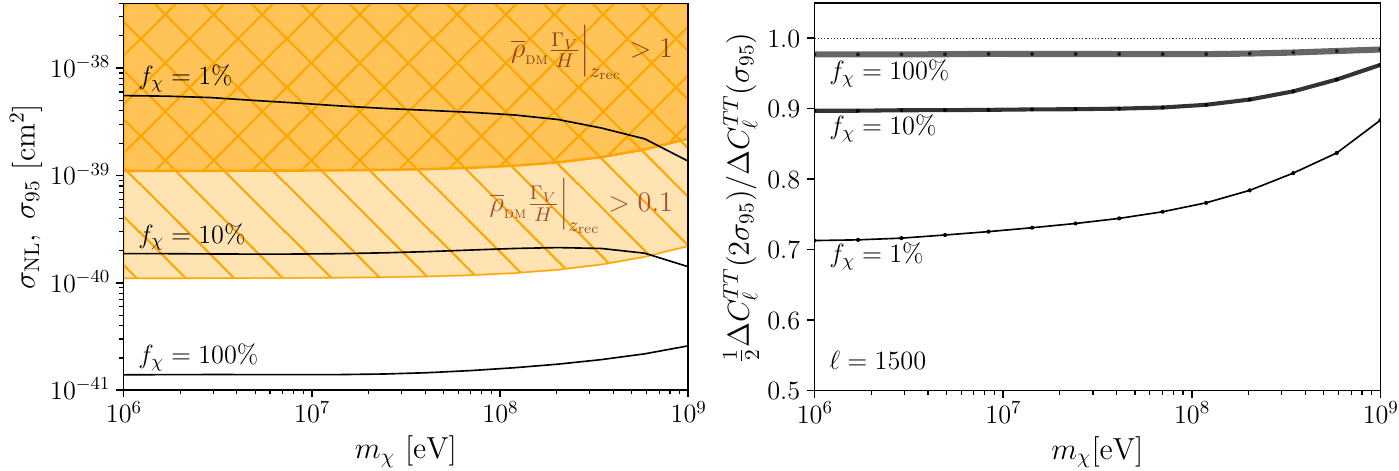} 
    \vspace{-0.385cm}
    \caption{Validity of the linear approximation for $\sigma_{\chi b}(v) \propto v^{-4}$ within the interacting-DM mass range, MeV$\leq m_\chi\leq$GeV, and for different fractions of interacting-DM, $f_\chi=\{1\%,10\%,100\%\}$. 
    \textbf{Left:} The shaded regions depict the range of cross-sections $\sigma_{\rm NL}$ for which the $\chi$-$b$ scattering efficiency around recombination ($z_{\rm rec}\approx 1100$) becomes nonlinear in $\sigma_{\chi b}$. Specifically, $\rhobar_{\rm DM}\Gamma_V/H$ becomes $\geq 0.1$ (light orange, \textbackslash-hatched) and $\geq 1$ (dark orange, $\times$-hatched) for cross-sections $\sigma_{\rm NL}$, where $\overline \rho_{\rm DM}$ is the total average DM density, and $H$ is the Hubble expansion rate. The black solid curves show $\sigma_{95}$, the 95\% C.L. CMB limits on cross-section [BG+18].   
    \textbf{Right:} The ratio $\Delta C_\ell^{TT}(2 \sigma_{95}) \big/ \big(2\Delta C_{\ell}^{TT}(\sigma_{95})\big)$ (at fixed $\ell=1500$) as a function of $m_\chi$, for three values of $f_\chi$. The closer the solid curves are to 1 (dotted horizontal line), the more linear is the dependence of $\Delta C_\ell^{TT}$ on $\sigma_{\chi b}$. We checked that the behavior of the curves does not change significantly for the $EE$ power spectrum, nor for any $\ell\gtrsim 1000$, and that the $f_{\chi} = 100\%$ curve for $\sigma_{\chi b}(v) \propto v^{-2}$ nearly overlaps with the corresponding curve for $\sigma_{\chi b}(v) \propto v^{-4}$ shown here.}\label{fig:DCl2}
\end{figure*}

We illustrate this point in the upper panels of Fig.~\ref{fig:DCl}, where we show the fractional changes in CMB temperature and $E$-mode polarization angular power spectra $\Delta C_\ell$, for cross-section $\sigma_{\chi b}(v) \propto v^{-4}$ set equal to the current 95\% CMB-anisotropy upper limits, and also to twice that value. We see that $\Delta C_\ell(2 \sigma_{\chi b}) \approx 2 \Delta C_\ell(\sigma_{\chi b})$ within a few percent relative difference, across all multipoles $\ell \leq 2500$, consistent with quasi-linearity of CMB power spectra in $\sigma_{\chi b}$. Note that the $\Delta C_\ell$ shown in Fig.~\ref{fig:DCl} are computed with the Boltzmann code \textsc{class} \cite{Blas:2011rf}, modified to account for DM-baryon scattering with the ansatz of DBK13, as described in Ref.~\cite{Becker:2020hzj}. As mentioned above, the DBK13 ansatz is fully nonlinear in $\sigma_{\chi b}$, thus provides a valid (even if approximate) prescription to test the linear dependence of an observable on $\sigma_{\chi b}$.

In contrast, if a small fraction of DM is interacting, observed CMB anisotropies can remain consistent with a relatively large DM-baryon cross section. Even if the effect of interactions on photons and baryons remains small in this case, the interaction may have non-perturbative effects on the subdominant interacting DM particle, by making it tightly coupled to baryons, i.e.~enforcing $\bs{V}_\chi \rightarrow \bs{V}_b$, and $V_{\chi b} \rightarrow 0$. As a consequence of the coupling of photons and baryons to interacting DM (through scattering and gravity) the observables may then depend nonlinearly on $\sigma_{\chi b}$, even if they are perturbatively affected by interactions. We illustrate this point in the lower panels of Fig.~\ref{fig:DCl}, where we show the change in angular power spectra for an interacting DM particle making up 1\% of the total DM abundance, and see that this change is not linear in $\sigma_{\chi b}$ near the 95\% upper limit from CMB anisotropies. A similar effect can be seen in Fig.~9 of BG+18, which shows that the fractional change in CMB power spectra is non-monotonic in $\sigma_{\chi b}$, and a fortiori nonlinear in this parameter, for a 0.3\% fraction of interacting DM.

In Fig.~\ref{fig:DCl2}, we further quantify where the current 95\% C.L. CMB upper limits, $\sigma_{95}$ \cite{Boddy:2018wzy}, for $n=-4$ lie with respect to the regimes of validity of the linear approximation. We do this for a range of DM masses, MeV$\leq m_\chi\leq$GeV, and the fraction of interacting DM $f_\chi=1, 10,$ and $100\%$. 
In the left panel, the hatched, shaded regions demarcate values of $\sigma_{\mathrm{NL}}$---the cross-sections for which the efficiency of $\chi$-$b$ momentum exchange (estimated by the ratio between the maximal scattering rate $\sim \rho_{\rm DM}\Gamma_V$ and the Hubble expansion rate $H$) becomes $\mathcal{O}(1)$ around recombination, implying that the scattering efficiency is nonlinear in $\sigma_{\chi b}$. We see that only the upper limits for $f_\chi=100\%$ lie safely within the linear regime (unhatched, unshaded), but for $f_\chi\lesssim 10\%$, the linear approximation can no longer be used to robustly compute the effect of $\chi$-$b$ scattering on the CMB power spectra.

We also demonstrate this point in the right panel of Fig.~\ref{fig:DCl2}, where the nonlinearity of $\Delta C_\ell$ increases (i.e., $\Delta C_\ell^{TT}(2 \sigma_{95}) \big/ \big(2\Delta C_{\ell}^{TT}(\sigma_{95})\big)$ deviates further from 1) monotonically as $f_\chi$ is decreased.

In both panels, for a given cross section and $f_\chi$, we see a trend with $m_\chi$ that the nonlinearity is more pronounced for lighter DM particles. This is because the smaller $m_\chi$ is, the larger the contribution of $T_\chi/m_\chi$ to the variance of relative thermal motions $(T/m)_{\chi b}$. And thus, for sufficiently light DM particles, the momentum-exchange coefficient $\Gamma_V(V_{\chi b}; (T/m)_{\chi b})$ becomes inherently nonlinear in $\sigma_{\chi b}$, because now in addition to its overall amplitude proportional to $\sigma_{\chi b}$, it also depends on $\sigma_{\chi b}$ through $(T/m)_{\chi b}$.

We cannot, a priori, expect CMB anisotropies to be quasi-linear in $\sigma_{\chi b}$ for cross sections scaling as $v^n$ with $n \geq -2$, since in these cases DM \emph{starts} tightly coupled to baryons, which is a non-perturbative effect. Still, we find that for $n = -2$, CMB anisotropy power spectra are very close to being linear in $\sigma_{\chi b}$ when all or most of the DM is interacting ($f_\chi\approx 100\%$), with $\Delta C_\ell(2 \sigma_{\chi b})/2 \Delta C_\ell(\sigma_{\chi b})$ within a few percent from unity across all relevant multipoles $\ell$ and masses $m_\chi$. This can be understood from the fact that, for $\sigma_{\chi b}$ saturating upper limits, the momentum-exchange rate $\overline{\rho}_b \Gamma_V$ becomes smaller than the expansion rate $H$ at very high redshift ($z \sim 10^7$, see e.g.~Fig.~5 of BG+18), i.e.~prior to the horizon entry of scales relevant to CMB anisotropies, thereby having only a nominal effect on those scales. 

For $\sigma_{\chi b}(v) \propto v^n$ with $n \geq 0$, however, CMB anisotropies are not quasilinear in the cross section. Luckily this is not an issue, since such cross sections are best constrained by probes sensitive to higher redshifts than CMB anisotropies (e.g.~Lyman-$\alpha$ forest \cite{Dvorkin:2013cea} or Milky-Way satellite counts \cite{Nadler:2019zrb}), at which point relative velocities are in any case negligible in comparison with their thermal counterparts \cite{Dvorkin:2013cea}.

In order to simplify the problem further, we focus on the regime in which the DM pressure term can be neglected. To estimate when this holds, we use the continuity equation for $\chi$'s density perturbation, $\dot{\delta}_\chi = - \theta_\chi$, and approximate time derivatives by a factor of the Hubble rate $H$. Only considering density perturbations in the pressure term, we thus have
\barr
\frac{\nabla^2 P_\chi/\overline{\rho}_\chi}{\dot{\theta}_\chi} &\sim& \frac{T_\chi}{m_\chi}(k/aH)^2  
\nonumber\\
&\leq& 0.015~ \frac{\textrm{MeV}}{m_\chi} (k~\textrm{Mpc})^2 \left(1 + z/z_{\rm eq}\right)^{-1},~
\earr
where $z \gg 1$ is the cosmological redshift, $z_{\rm eq} \approx 3.3\times 10^3$ the redshift at matter-radiation equality, wavenumbers $k$ are comoving, and we used $T_\chi \leq T_b \leq 2.73 (1+z)~$K, which holds when the dissipation of relative velocities can be neglected. We thus see that, for $m_\chi \gtrsim $ MeV, one may safely neglect the DM pressure term in the momentum equations (see also Ref.~\cite{Becker:2020hzj} for a similar conclusion). Note that this is a conservative lower limit on the DM mass for which pressure may be neglected, since in most cases the DM temperature falls below that of baryons well before matter-radiation equality.

\subsection{Linear-response expressions}\label{sec:Linear-response}

Having defined the regime of ``weak interactions", we now expand all fluid variables to first-order in the cross section, i.e.~write them as $X = X^{(0)} + X^{(1)}$, where the zeroth-order term $X^{(0)} = X^{\rm std}$ is independent of $\sigma_{\chi b}$ and corresponds to the standard CDM-only scenario, the second term $X^{(1)}$ scales linearly with $\sigma_{\chi b}$, and we neglect terms of order $\sigma_{\chi b}^2$ or higher. This expansion applies not only to the $\chi$ and baryon variables explicitly mentioned so far, e.g.~$\bs{V}_b = \bs{V}_b^{(0)} + \bs{V}_b^{(1)}$, but also to all the other fluid and metric variables, which are coupled to baryons and $\chi$ through gravity and Thomson scattering. Note that we do not attempt to identify a specific cross section $\sigma_*$ relative to which $\sigma_{\chi b}$ must be small for this expansion to be accurate; instead, we simply use the fact that fluid and metric variables must be \emph{analytic functions} of $\sigma_{\chi b}(v)$ (at any given velocity $v$), hence can be Taylor-expanded in this parameter for sufficiently small cross sections. 

Since the momentum-exchange rate is already linear in $\sigma_{\chi b}$, to solve for the first-order fluid variables we only need to evaluate $\Gamma_V$ at the zero-th order values of $V_{\chi b}$ and $(T/m)_{\chi b}$, the latter being simply $T_b^{(0)}/m_b$, since $T_\chi^{(0)} = 0$. Since, moreover, relative baryon temperature perturbations are small, given that baryons are thermally coupled to photons down to $z \sim 200$, we may simply evaluate $\Gamma_V$ at the standard mean baryon temperature $\overline{T}_b^{(0)} = \overline{T}_b^{\rm std}$. 

Even though our results can easily be generalized to e.g.~the matter power spectrum used for the Lyman-$\alpha$ forest analysis, since CMB anisotropies are most constraining for cross sections $\sigma_{\chi b}(v) \propto v^{-4}$, for definiteness we focus the discussion on this specific observable in what follows.

Let us now denote by $\Theta_{\ell m}$ the spherical harmonic amplitudes of the CMB temperature anisotropy. We also expand it to first-order in the cross section: $\Theta_{\ell m} = \Theta^{(0)}_{\ell m} + \Theta^{(1)}_{\ell m}$. The zeroth-order (i.e.~standard) piece is linear in the primordial curvature perturbation $\zeta(\bs{x})$, and can be written in the general form
\beq
\Theta_{\ell m}^{(0)} = \Theta_{\ell m}^{\rm std} = \int d^3 x ~\mathcal{T}_{\ell m}(\bs{x}) \zeta(\bs{x}), \label{eq:Theta0}
\eeq
where $\mathcal{T}_{\ell m}(\bs{x})$ is the standard linear transfer function, obtained with no DM-baryon scattering -- recalling that, in this context, ``linear" refers to the the scaling with primordial initial conditions.

In the regime where the DM pressure force may be neglected, the only effect of DM-baryon scattering on fluid variables relevant to CMB anisotropies is through the momentum-exchange terms in Eqs.~\eqref{eq:dotVchi}, \eqref{eq:dotVb}, which are both proportional to $\bs{S} \equiv \Gamma_V \bs{V}_{\chi b}$. At first order in $\sigma_{\chi b}$, all the perturbed fluid variables are thus proportional to the first-order field
\beq\label{eq:S}
\bs{S}^{(1)}(t, \bs{x})= \Gamma_V\left(V_{\chi b}^{(0)}(t, \bs{x}); \overline{T}_b^{(0)}(t)/m_b\right) \bs{V}_{\chi b}^{(0)}(t, \bs{x}),
\eeq
where we have written explicitly all the temporal and spatial dependencies. 

Since $\Theta$ is sourced linearly by fluid variables, which at first order are sourced linearly by the field $\bs{S}^{(1)}$, we also have, very generally,
\beq
\Theta_{\ell m}^{(1)} = \iint dt ~d^3 x ~ \bs{\mathcal{G}}_{\ell m}(t, \bs{x}) \cdot \bs{S}^{(1)}(t, \bs{x}), \label{eq:Theta1}
\eeq
where $\bs{\mathcal{G}}_{\ell m}(t, \bs{x})$ is a vector Green's function.

Note that for any practical computation, one would rather work in Fourier space, but these real-space expressions turn out to be most convenient for the purpose of our formal proof. Analogous expressions can be written for the $E$-mode polarization field $E_{\ell m} = E_{\ell m}^{(0)} + E_{\ell m}^{(1)}$. While there is no zeroth-order $B$ mode for scalar initial conditions, note that DM-baryon scattering should source a $B$-mode at first order, $B_{\ell m}^{(1)} \neq 0$, also linear in $\bs{S}^{(1)}(t, \bs{x})$.

The CMB temperature power spectrum $C_\ell^{TT}$ is defined through $
\langle \Theta_{\ell m} \Theta_{\ell' m'}^* \rangle = \delta_{\ell \ell'} \delta_{m m'} C_{\ell}^{TT}$, which implies that, at linear order in $\sigma_{\chi b}$, 
\barr
C_\ell^{TT} &=& C_\ell^{TT,\rm std} + \Delta C_\ell^{TT}, \\
 \Delta C_{\ell}^{TT} &=& 2\textrm{Re}\left( \langle \Theta_{\ell m}^{*(0)}\Theta_{\ell m}^{(1)} \rangle\right), 
\earr
where $m$ can take any value in $[-\ell,\ell]$. Analogous expressions can be written for the polarization power spectrum $C_\ell^{EE}$ and the temperature-polarization cross power spectrum $C_{\ell}^{TE}$. From Eqs.~\eqref{eq:Theta0} and \eqref{eq:Theta1}, and their analogs for polarization, we see that, at linear order in $\sigma_{\chi b}$, the perturbations $\Delta C_\ell^{TT}, \Delta C_\ell^{TE}, \Delta C_\ell^{EE}$ are entirely determined by the cross-correlation of the initial curvature perturbation $\zeta$ with the source function $\bs S^{(1)}$; we define this \textit{vector} cross-correlation function as:
\beq
\bs{\Xi}(t, \bs{r}) \equiv \langle \zeta(\bs{x}) \bs{S}^{(1)}(t, \bs{x} + \bs{r})\rangle = \langle \zeta(\bs{0}) \bs{S}^{(1)}(t, \bs{r})\rangle, 
\eeq
where we used statistical homogeneity in the second equality. Explicitly, we have, for instance,
\barr
\Delta C_\ell^{TT} = 2 \textrm{Re} \Bigg{(} \iiint dt~ d^3x ~d^3r ~\mathcal{T}_{\ell m}^*(\bs{x}) \nonumber\\
\times \bs{\mathcal{G}}_{\ell m}(\bs{x}+\bs{r})\cdot \bs{\Xi}(t, \bs{r}) \Bigg{)},
\earr
with similar expressions for $\Delta C_\ell^{EE}$ and $\Delta C_\ell^{TE}$.

\subsection{Calculation of $\bs{\Xi}(t, \bs{r})$}\label{sec:calcXi}

We now compute the two-point correlation function $\bs{\Xi}(\bs{r})$, where from now on we no longer write the time dependence explicitly. To simplify the notation, we define $\zeta_{\bs 0} \equiv \zeta(\bs{0})$, $\bs{V}_{\bs{r}} \equiv \bs{V}_{\chi b}^{(0)}(\bs{r})$. We are therefore interested in computing
\beq \label{eq:Xi}
\bs{\Xi}(\bs{r}) = \langle \zeta_{\bs 0} \Gamma_V(V_{\bs{r}}) \bs{V}_{\bs{r}}\rangle,
\eeq
where we have replaced $\bs S^{(1)}$ using Eq.~\eqref{eq:S}.
The average in Eq.~\eqref{eq:Xi} is to be taken over the correlated Gaussian distribution of $\zeta_{\bs 0}$ and $\bs{V}_{\bs{r}}$. We may rewrite it as follows:
\barr
\bs{\Xi}(\bs{r}) = \Big{\langle} \langle  \zeta_{\bs 0}|\bs{V}_{\bs{r}}\rangle \Gamma_V(V_{\bs{r}}) \bs{V}_{\bs{r}} \Big{\rangle}, \label{eq:nested-av}
\earr
where the inner average $\langle \cdots | \bs{V}_{\bs{r}} \rangle$ is over the constrained Gaussian distribution of $\zeta_0$ at fixed $\bs{V}_{\bs r}$ and the outer average is over the unconstrained 3-dimensional Gaussian distribution of $\bs V_{\bs r}$.   

We show in Appendix \ref{app:constrained} that the constrained average of $\zeta_0$ at fixed $\bs{V}_{\bs{r}}$ takes the form
\barr \label{eq:1st constrained avg}
\langle \zeta_{\bs 0} | \bs{V}_{\bs{r}} \rangle = \frac{3}{\langle V_{\bs r}^2\rangle } \langle \zeta_0 \bs{V}_{\bs{r}} \rangle \cdot \bs{V}_{\bs{r}}.
\earr
To be clear, the $\bs{V}_{\bs r}$ appearing in the conditional average on the left-hand-side and outside the brackets on the right-hand-side both represent the same specific value of the relative velocity field at position $\bs{r}$, while the $\bs{V}_{\bs r}$ appearing inside the brackets on the right-hand-side is a dummy variable to be averaged over (as is $\zeta_0)$.

Inserting Eq.~\eqref{eq:1st constrained avg} into Eq.~\eqref{eq:nested-av}, we find that the $i$-th component of $\bs{\Xi}(\bs{r})$ is
\barr
\Xi^i(\bs{r}) &=& \Big{\langle} \langle \zeta_{\bs 0} | \bs{V}_{\bs{r}} \rangle \Gamma_V(V_{\bs{r}}) V_{\bs{r}}^i \Big{\rangle}\nonumber\\
&=& \frac{3}{\langle V_{\bs r}^2 \rangle} \langle \zeta_0 V_{\bs{r}}^j\rangle  \langle V_{\bs r}^j \Gamma_V(V_{\bs{r}}) V_{\bs r}^i  \rangle,
\earr
where the last average is now to be taken over the Gaussian and isotropic distribution of $\bs{V}_{\bs{r}}$, which is in fact independent of $\bs{r}$. In the last average, we perform the angular average over $\hat{\bs V}_{\bs r}$ first, giving $\langle \hat{V}_{\bs{r}}^i \hat{V}_{\bs{r}}^j \rangle = \frac13 \delta_{ij}$. This implies the following explicit expression for $\bs{\Xi}(t, \bs{r})$:
\beq
\bs{\Xi}(t, \bs{r}) = \frac{\langle V^2 \Gamma_V(V) \rangle_t}{\langle V^2 \rangle_t} ~ \langle \zeta_0 \bs{V}_{\bs{r}}\rangle_t, \label{eq:final Xi}
\eeq
where we have made explicit that the averages of functions of $\bs{V}_{\bs{r}}$ alone are independent of $\bs{r}$, and highlighted that all these averages are time dependent. We see that the cross-correlation $\bs{\Xi}(t, \bs{r})$ has a universal $\bs{r}$ dependence, which is that of the correlation function $\langle \zeta_0 \bs{V}_{\bs r} \rangle$.

\subsection{A powerful result}\label{sec:powerfulResult}

Equation \eqref{eq:final Xi} proves that the correlation function $\bs{\Xi}(t,\bs{r})$, and therefore the first-order perturbations to CMB power spectra $\Delta C_\ell^{TT,EE,TE}$, depend on the function $\Gamma_V(V)$ only through its time-dependent second moment $\langle V^2 \Gamma_V(V) \rangle_t$, taken over the Gaussian distribution of \emph{standard} (no DM-baryon scattering) relative velocities $\bs{V}_{\chi b}^{(0)}$ at time $t$. 

A corollary of this finding is that any two functions $\Gamma_V(V)$ and $\widetilde{\Gamma}_V(V)$ with the same second moment at all times would result in identical first-order perturbations to CMB power spectra. This result can be harnessed to simplify calculations: to obtain the correct CMB power spectra at linear order in $\sigma_{\chi b}$, it suffices to compute them with the simplest possible alternative momentum-exchange coefficient $\widetilde{\Gamma}_V(V)$, as long as its second moment $\langle V^2 \widetilde{\Gamma}_V(V) \rangle$ matches that of the original $\Gamma_V$ at all times. The simplest possible $\widetilde\Gamma_V$ obeying this constraint is the velocity-independent coefficient
\beq\label{eq:tildeGamma}
\widetilde{\Gamma}_V(t) = \frac{\langle V^2 \Gamma_V(V) \rangle_t}{\langle V^2\rangle_t}.
\eeq 
This is the \emph{unique} velocity-independent coefficient with the same second moment as $\Gamma_V(V)$.

This is a very powerful result: we have demonstrated that, in order to recover the exact CMB power spectra at linear order in $\sigma_{\chi b}$, it suffices to solve the Einstein-Boltzmann fluid equations with the specific \emph{velocity-independent} momentum-exchange coefficient $\widetilde{\Gamma}_V$ given in Eq.~\eqref{eq:tildeGamma}, instead of the exact velocity-dependent $\Gamma_V(V)$. This result allows for a tremendous simplification of the problem. Let $\widetilde{\bs{V}}_\chi, \widetilde{\bs{V}}_b$, etc... be the fields obtained by solving the momentum equations (\ref{eq:dotVchi}, \ref{eq:dotVb}) with $\Gamma_V \rightarrow \widetilde{\Gamma}_V$. The velocity independence of $\widetilde \Gamma_V$ makes the momentum equations \emph{linear} in the velocity fields $\widetilde{\bs V}_{\chi}, \widetilde{\bs V}_b$. Moreover, it implies that $\widetilde{\Gamma}_V$ is homogenous. As a consequence, for scalar adiabatic initial conditions, the velocity fields $\widetilde{\bs{V}}_\chi, \widetilde{\bs{V}}_b$ are indeed curl-free and fully described by their divergences $\widetilde{\theta}_\chi, \widetilde{\theta}_b$, which do satisfy Eqs.~(\ref{eq:dottheta_chi}, \ref{eq:dottheta_b}).

In summary, we have derived, from first principles, the \emph{unique} velocity-independent substitution for the momentum-exchange coefficient $\Gamma_V$ which allows to recover CMB power spectra \emph{exactly} at linear order in $\sigma_{\chi b}$. This significantly improves upon the ansatzes proposed by DBK13 and BG+18, whose accuracy was unknown, as they did not arise from a first-principle calculation.

\subsection{A serendipitous coincidence} \label{sec:coincidence}

We have just shown that it suffices to replace $\Gamma_V(V)$ in the fluid equations with $\widetilde{\Gamma}_V$ from Eq.~\eqref{eq:tildeGamma} to get the correct CMB power spectra at linear order in $\sigma_{\chi b}$. We now present the unexpected finding that $\widetilde{\Gamma}_V$ happens to be precisely equal to the ansatz proposed by DBK13, given in Eq.~\eqref{eq:Dvorkin}.

To show this, we need an explicit expression for $\Gamma_V(V, T/m)$, for an arbitrary value of the bulk relative velocity $V$ and of its variance per axis $T/m$. A general expression based on the microphysics of elastic scattering can be obtained from combining Eqs.~(55) and (21) of Ref.~\cite{Ali-Haimoud:2018dvo}:
\barr\label{eq:microphGamma}
\Gamma_V(V, T/m) 
= \frac1{M} \int d^3 v ~  \frac{\bs{v} \cdot \bs{V}}{V^2} v\, \sigma_{\chi b}(v) ~G\left(\bs{v}| \bs{V}, T/m\right),~\label{eq:GammaV}
\earr
where $M \equiv m_b + m_\chi$ and $G(\bs{v}|\bs{V}, T/m) $ is the Gaussian distribution of the \emph{local} particle relative velocities $\bs v$, with mean $\bs{V}$ and variance per axis $T/m$. The second moment of Eq.~\eqref{eq:microphGamma} is then

\barr
\langle V^2 \Gamma_V(V, T/m) \rangle 
= \frac1{M} \iint d^3 V d^3 v ~(\bs{v} \cdot \bs{V}) ~v ~\sigma_{\chi b}(v)\nonumber\\
 ~~~\times  G(\bs{V}| \bs{0}, \langle V^2 \rangle/3) G\left(\bs{v}|\bs{V}, T/m\right),~ \label{eq:V2GammaV}
\earr
where $G(\bs{V}|\bs{0}, \langle V^2 \rangle/3)$ is the Gaussian distribution of \emph{cosmological} bulk relative velocities $\bs{V}$, with zero mean and variance per axis $\langle V^2\rangle/3$. We may rewrite the product of the two Gaussians as
\barr
&&G(\bs{V}|\bs{0}, \langle V^2 \rangle/3) G\left(\bs{v}|\bs{V}, T/m\right)\nonumber\\
 &=& G(\bs{v}| \bs{0}, T/m + \langle V^2 \rangle/3)  G\left(\bs{V}| \kappa ~ \bs{v}, \kappa~ T/m \right),~~~~~\\
\kappa &\equiv& \frac{\langle V^2 \rangle}{3 T/m + \langle V^2 \rangle }.~~~~~~
\earr
This expression isolates the $\bs V$-dependence and can be obtained by simply completing the squares in the exponents of the Gaussians. Substituting this result into Eq.~\eqref{eq:V2GammaV}, we can now integrate over $\bs{V}$, and obtain
\barr \label{eq:tildeGammaVintegrated}
\widetilde{\Gamma}_V &=& \frac{\langle V^2 \Gamma_V(V, T/m) \rangle}{\langle V^2 \rangle} \nonumber\\
&=&  \frac1{M} \int d^3 v ~ v^3 \sigma_{\chi b}(v) ~ \frac{G(\bs{v}|\bs{0}, T/m + \langle V^2\rangle /3)}{3 T/m + \langle V^2\rangle}.~~~~ \label{eq:V2Gamma-almost}
\earr
Let us now compute $\Gamma_V(0, \sigma_{1\rm D}^2)$, for a vanishing mean of local particle relative velocities $(V = 0)$, and arbitrary variance per axis $\sigma_{1 \rm D}^2$. We start by Taylor-expanding the Gaussian distribution of relative velocities to first order in $\bs{V}$:
\barr
G(\bs{v}|\bs{V}, \sigma_{1 \rm D}^2) &=& G(\bs{v}|\bs{0}, \sigma_{1 \rm D}^2)\nonumber\\
&&\times \Big{[} 1 + (\bs{v}\cdot \bs{V})/\sigma_{1 \rm D}^2 + \mathcal{O}(V^2) \Big{]}. ~~~~~
\earr
Inserting this expression into Eq.~\eqref{eq:GammaV}, we see that the first term integrates to zero, and the dependence on $V$ cancels out in the second one, so that
\barr
\Gamma_V\left(0, \sigma_{1 \rm D}^2\right) = \frac1{M}  \int d^3v~ v^3 \sigma_{\chi b}(v)~ \frac{G(\bs{v}|\bs{0}, \sigma_{1 \rm D}^2)}{3\sigma_{1 \rm D}^2}. \label{eq:Gamma_V(0, sigma1D)}
\earr
Comparing Eq.~\eqref{eq:Gamma_V(0, sigma1D)} with Eq.~\eqref{eq:V2Gamma-almost}, we see that, 
\beq
\widetilde{\Gamma}_V = \Gamma_V\left(0,  T/m + \langle V^2\rangle/3\right),
\eeq
which, as advertized, is the ansatz proposed by DBK13. 

In conclusion, we have shown that the ansatz taken by DBK13 is, coincidentally, precisely the correct velocity-independent coefficient needed to get exact CMB power spectra at linear order in the cross section. 

\subsection{Generalization to multiple baryon scatterers}\label{sec:multiple-scat}

Let us end by considering the general case where (part of) the DM scatters with multiple baryonic species (Hydrogen and Helium atoms and/or nuclei and free electrons). The frequent scattering of different baryonic species among themselves implies that they are tightly coupled, hence all have the same bulk velocity $\bs{V}_b$. Considering again the regime where that DM pressure is negligible, the momentum-exchange equations become
\barr
\dot{\bs{V}}_\chi &=& \dot{\bs{V}}_\chi\big{|}_{\rm std}  - \overline{\rho}_b \sum_s \frac{\overline{\rho}_s}{\overline{\rho}_b} \Gamma_V^s \bs{V}_{\chi b} \label{eq:dotVchi-mult}\\
\dot{\bs{V}}_b &=& \dot{\bs{V}}_b\big{|}_{\rm std} + \overline{\rho}_\chi \sum_s \frac{\overline{\rho}_s}{\overline{\rho}_b}~ \Gamma_V^s \bs{V}_{\chi b} \label{eq:dotVb-mult},
\earr
where the index $s$ labels the different baryonic scatterers, and the momentum-exchange coefficients are each computed with the relevant cross section $\sigma_{\chi s}(v)$. These equations take exactly the same form as Eqs.~(\ref{eq:dotVchi},\ref{eq:dotVb}), with $\Gamma_V(V)$ replaced with the mass-density-weighted momentum-exchange coefficient $\sum_s (\overline{\rho}_s/\overline{\rho}_b) \Gamma_V^s(V)$. The reasoning leading to the exact velocity-independent substitution Eq.~\eqref{eq:tildeGamma} is identical: to get the exact CMB power spectra at linear order in \emph{all} the cross sections $\sigma_{\chi s}$, it suffices to replace $\Gamma_V$ with $\langle V^2 \Gamma_V(V)\rangle/\langle V^2 \rangle$, which amounts to substituting each individual $\Gamma_V^s$ with $\langle V^2 \Gamma_V^s(V)\rangle/\langle V^2 \rangle$. One can also arrive at this result by simply pointing out that, in the regime where the effect of DM-baryon scattering is linear in \emph{each} $\sigma_{\chi s}$, the total effect of scattering with multiple species is obtained by linear superposition of the individual effects of each scatterer.

\section{Conclusions and outlook} \label{sec:conclusion}
Accurately describing the effect of DM-baryon interactions on cosmological observables has long been a technical challenge, due to the dependence of the momentum-exchange coefficient $\Gamma_V(V_{\chi b})$ on the local DM-baryon mean relative velocities $V_{\chi b}$, which in turn implies a nonlinear momentum-exchange rate. In order to include DM-baryon scattering within linear Boltzmann codes, the standard approach has been to substitute $\Gamma_V(V_{\chi b})$ with a phenomenological velocity-independent coefficient $\Gamma_V^{\rm pheno}$ \cite{Dvorkin:2013cea, Boddy:2018wzy}. However, such ansatzes are not unique, and were never derived from first principles. It was therefore not clear whether such phenomenological substitutions may be accurate beyond the order-unity level. 

In this work, we derive, for the first time, a rigorous approximation scheme to deal with DM-baryon scattering, by constructing an expansion in powers of the DM-baryon cross section $\sigma_{\chi b}$. We demonstrate, that, at linear order in $\sigma_{\chi b}$, CMB anisotropy power spectra (or in general, the power spectra of any linear-cosmology observable) only depend on the coefficient $\Gamma_V(V_{\chi b})$ through its second moment, taken over the Gaussian distribution of unperturbed mean relative velocities. This is a powerful result, as it implies that it suffices to solve the Boltzmann-fluid equations with the simple, velocity-independent coefficient $\widetilde{\Gamma}_V = \langle V_{\chi b}^2 \Gamma_V(V_{\chi b})\rangle/\langle V_{\chi b}^2\rangle$, to recover all CMB anisotropy power spectra \emph{exactly} at first-order in the cross section. 

Moreover, we prove that, for any velocity-dependent cross section $\sigma_{\chi b}(v)$, the coefficient $\widetilde{\Gamma}_V$ happens to be precisely equal to the most commonly used phenomenological coefficient $\Gamma_V^{\rm pheno}$ proposed in Ref.~\cite{Dvorkin:2013cea}. This serendipitous coincidence places current DM-baryon scattering upper limits on a much firmer footing: whenever an observable is within the regime of linear dependence on $\sigma_{\chi b}$, the standard phenomenological coefficient happens to lead to the correct power spectra, up to corrections quadratic in $\sigma_{\chi b}$. We argue that this is the case for CMB anisotropy power spectra limits when \emph{all} the DM is interacting with baryons with a cross section $\sigma_{\chi b}(v) \propto v^{-2}, v^{-4}$. It is not the case when only a small fraction of DM interacts with baryons, as it may be affected non-perturbatively in $\sigma_{\chi b}$ \cite{Boddy:2018wzy}. This means that current upper limits in this regime should be understood to be accurate only at the order-unity level. Still, the fact that one can find an exact solution in some well-defined regime is significant progress.

It is worth emphasizing that the perturbed CMB temperature and $E$-mode polarization fields $\widetilde{T}$ and $\widetilde{E}$ obtained with the coefficient $\widetilde{\Gamma}_V$ \emph{are not equal} to the exact $T$ and $E$ fields that would result from solving the fluid equations with the velocity-dependent momentum-exchange coefficient $\Gamma_V(V_{\chi b})$. It is only their auto- and cross- \emph{power spectra} which are identical, at linear order in $\sigma_{\chi b}$. Importantly, other statistical properties need not match. In particular, while $\widetilde{T}$ and $\widetilde{E}$ are Gaussian, the exact $T$ and $E$ fields are non-Gaussian \cite{Dvorkin:2013cea}, due to their nonlinear dependence on initial conditions, stemming from the nonlinear momentum-exchange rate. Given that the latter only contains odd powers of relative velocities, we expect the lowest-order non-Gaussian correlation functions to be trispectra or connected 4-point functions, such as $\langle TTTT \rangle_c, \langle TTTE\rangle_c$, etc. Although very different in physical origin, this signature is qualitatively similar to that of inhomogeneously accreting primordial black holes (PBHs) \cite{Jensen:2021mik, Jensen:2022zzt}. 
As in the case of PBHs, we expect the trispectra to be more constraining probes of DM-baryon interactions than power spectra. Furthermore, the trispectrum formalism is significantly more complex and beyond the scope of this paper, but it is the subject of our work currently in progress.

Excitingly, while the $B$-mode polarization field computed with $\widetilde{\Gamma}_V$ vanishes identically for scalar initial conditions, we expect that the exact $B$-mode field not only does not vanish, but is also non-Gaussian. This implies that DM-baryon scattering should source $B$-mode trispectra, such as $\langle T T T B \rangle_c, \langle T T E B \rangle_c$, etc... Note that, while the cross-spectra $\langle TB\rangle$ and $\langle EB \rangle$ always vanish in the absence of parity-violating physics, higher-order correlation functions containing a $B$ mode need not vanish \cite{Meerburg:2016ecv}. Such $B$-mode trispectra are particularly interesting due to the low cosmic-variance noise of $B$-mode polarization \cite{Meerburg:2016ecv}. Building on our perturbative expansion framework, we will explore these promising new probes of DM-baryon scattering in future work.

\section*{Acknowledgments}

We thank Kimberly Boddy and Vera Gluscevic for useful conversations, and for sharing the data of BG+18. We thank Marc Kamionkowski and Julien Lesgourgues for encouragements. YAH is grateful to the USC Department of Physics and Astronomy for hosting him during his sabbatical, when part of this work was completed, and acknowledges support from the CIFAR-Azrieli Global Scholars program. YAH and SSG are supported by NASA grant 80NSSC20K0532. SSG is also supported by the AAUW American Fellowship and the NYU James Arthur Graduate Associate Fellowship. TLS is supported by NSF Grants No.~2009377 and No.~2308173 and thanks the NYU CCPP where part of his contribution to this work was completed. 

\appendix

\section{Constrained average $\langle \zeta_0 |\bs{V}_{\bs r} \rangle $}\label{app:constrained}

The 4-dimensional joint Gaussian distribution of $\zeta_0$ and $\bs{V}_{\bs r}$ takes the form
\barr
\mathcal{P}_{4 \rm D}(\zeta_0, \bs{V}_{\bs{r}}) &=& \frac1{(2 \pi)^2\sqrt{\det \bs{C}}} ~e^{-\frac12 \bs{X} \cdot \bs{C}^{-1} \cdot \bs{X}^{\rm T}},~~~\\
\bs{X} &\equiv& [\zeta_0, V_{\bs r}^1, V_{\bs r}^2, V_{\bs r}^3], 
\earr
where the 4 by 4 covariance matrix is given by
\beq
\bs{C} = \left[\begin{matrix} \langle \zeta_0^2\rangle & \langle \zeta_0 V_{\bs r}^1 \rangle & \langle \zeta_0 V_{\bs r}^2 \rangle & \langle \zeta_0 V_{\bs r}^3 \rangle \\
\langle \zeta_0 V_{\bs r}^1 \rangle & \langle V^2 \rangle/3 & 0 & 0  \\
\langle \zeta_0 V_{\bs r}^2 \rangle & 0 & \langle V^2 \rangle/3& 0 \\
\langle \zeta_0 V_{\bs r}^3 \rangle & 0  & 0  & \langle V^2 \rangle/3 
\end{matrix}\right],
\eeq
where we used the fact that $\langle V_{\bs{r}}^i V_{\bs r}^j \rangle = \delta_{ij} \langle V_{\bs r}^2 \rangle/3 = \langle V^2 \rangle/3$, since the (unconstrained) distribution of relative velocities is isotropic and independent of $\bs{r}$. The inverse of this matrix can be computed explicitly:
\barr
\bs{C}^{-1} &=& \frac1{\Delta}\left[\begin{matrix} \langle V^2\rangle/3 & -\langle \zeta_0 V_{\bs r}^1 \rangle & -\langle \zeta_0 V_{\bs r}^2 \rangle & -\langle \zeta_0 V_{\bs r}^3 \rangle \\
-\langle \zeta_0 V_{\bs r}^1 \rangle & M_{11} & M_{12} & M_{13}  \\
-\langle \zeta_0 V_{\bs r}^2 \rangle & M_{21} & M_{22}& M_{23}\\
-\langle \zeta_0 V_{\bs r}^3 \rangle & M_{31}  & M_{32}  & M_{33} 
\end{matrix}\right],~~~~~~\\
\Delta &\equiv& \langle \zeta_0^2 \rangle \langle V^2 \rangle/3 - \sum_i\langle \zeta_0 V_{\bs r}^i \rangle^2,\\
M_{ij} &\equiv& \frac3{\langle V^2\rangle}\left(\Delta \delta_{ij} + \langle \zeta_0 V_{\bs r}^i \rangle  \langle \zeta_0 V_{\bs r}^j \rangle\right).
\earr
We therefore have
\barr
\bs{X} \cdot \bs{C}^{-1} \cdot \bs{X}^{\rm T} = \frac{\langle V^2\rangle \zeta_0^2 }{3 \Delta} - \frac2{\Delta} \langle \zeta_0 \bs{V}_{\bs r}\rangle \cdot (\zeta_0 \bs{V}_{\bs r})\nonumber\\
+ \frac{3 \bs{V}_{\bs r}^2}{\langle V^2 \rangle} + \frac3{\Delta \langle V^2 \rangle} \left(\langle \zeta_0 \bs{V}_{\bs r} \rangle \cdot \bs{V}_{\bs r}\right)^2\nonumber\\
= \frac{3 \bs{V}_{\bs r}^2}{\langle V^2 \rangle} + \frac{\langle V^2 \rangle}{3 \Delta }\left(\zeta_0 - 3 \frac{\langle \zeta_0 \bs{V}_{\bs{r}}\rangle \cdot \bs{V}_{\bs r}}{\langle V^2 \rangle}\right)^2.
\earr
From the last expression, we see that we may factorize the 4D joint Gaussian distribution for $\zeta_0$ and $\bs{V}_{\bs r}$ as the product of a 3D (unconstrained) Gaussian distribution for $\bs{V}_{\bs r}$, with zero mean and variance $\langle V^2 \rangle/3$ per axis, and a constrained 1D Gaussian distribution for $\zeta_0$ at fixed $\bs{V}_{\bs r}$, with variance $3 \Delta/\langle V^2 \rangle$ and mean $\langle \zeta_0 |\bs{V}_{\bs r} \rangle = 3 \langle \zeta_0 \bs{V}_{\bs{r}}\rangle \cdot \bs{V}_{\bs r}/\langle V^2 \rangle$.

\newpage

\bibliography{main_v2.bib}
\end{document}